\documentclass{article}

\PassOptionsToPackage{square,numbers}{natbib}
\usepackage[final]{neurips_2021}

\usepackage[utf8]{inputenc} % allow utf-8 input
\usepackage[T1]{fontenc}  % use 8-bit T1 fonts
\usepackage{hyperref}    % hyperlinks
\usepackage{url}      % simple URL typesetting
\usepackage{booktabs}    % professional-quality tables
\usepackage{amsfonts}    % blackboard math symbols
\usepackage{nicefrac}    % compact symbols for 1/2, etc.
\usepackage{microtype}   % microtypography
\usepackage{xcolor}     % colors
\usepackage{graphicx}    % Graphics
\usepackage{amsmath}    % \AA
\usepackage{indentfirst}
\usepackage{color, colortbl}
\definecolor{Gray}{gray}{0.9}

\title{Learning Small Molecule Energies and Interatomic Forces with an Equivariant Transformer on the ANI-1x Dataset}

\author{
 Bryce Hedelius\\
 Department of Physics and Astronomy\\
 Brigham Young University\\
 \AND
 Fabian B. Fuchs\\
 A2I Lab, Engineering Department\\
 Oxford University\\
 \AND
 Dennis Della Corte\\
 Department of Physics and Astronomy\\
 Brigham Young University\\
 dennis.dellacorte@byu.edu\\
}

\begin{document}

\maketitle

\begin{abstract}
	Accurate predictions of interatomic energies and forces are essential for high quality molecular dynamic simulations (MD). Machine learning algorithms can be used to overcome limitations of classical MD by predicting ab initio quality energies and forces. SE(3)-equivariant neural network allow reasoning over spatial relationships and exploiting the rotational and translational symmetries. One such algorithm is the SE(3)-Transformer, which we adapt for the ANI-1x dataset. Our early experimental results indicate through ablation studies that deeper networks---with additional SE(3)-Transformer layers---could reach necessary accuracies to allow effective integration with MD. However, faster implementations of the SE(3)-Transformer will be required, such as the recently published accelerated version by Milesi. 
\end{abstract}

\section{Introduction}

The accurate assignment of energies and forces to atoms of small molecules is essential to many branches of chemistry and material science.\cite{burke2012} Multiple computational methods have been developed to solve this many-body quantum mechanical problem, such as Density Functional Theory (DFT) \cite{engel2013} and Coupled Cluster (CC) theory \cite{bartlett2007}. 
More recently, machine learning methods have been trained on datasets obtained from DFT or CC and are shown to be able to predict energies within errors comparable to the accuracy of DFT.\cite{batzner2021} One well-known algorithm is ANI \cite{smith2017-network}, for which different training datasets are available \cite{smith2017-dataset, smith2018} that have been created with support of unsupervised training \cite{smith2018}. ANI is still being developed and was recently extended \cite{devereux2020} beyond the initial four atom types and is also available as a PyTorch implementation \cite{gao2020}. The ANI network computes atomic environment vectors (AEV) for each atom in a system and passes the vectors through a dedicated fully connected neural network corresponding to the atom type. 

 A major limitation of ANI is the necessity to train dedicated neural networks for each atom type and the usage of handcrafted features like AEV. Graph neural networks provide a more general approach for the prediction of atomic energies and forces. A variety of neural networks has been designed to leverage symmetries of tasks on molecular graphs, with equivariance and invariance as their foundational principles.\cite{cohen2016, cohen2018theory} Equivariance for translation and rotation in 3D (symmetries described by the SE(3) group) were first investigated by Cohen et al. \cite{cohen2018spherical} , Esteves et al. \cite{exteves2017}, and Kondor et al. \cite{kondor2018} and were transferred to 3D graphs of molecules by Anderson et al. \cite{anderson2019}, Batzner et al. \cite{batzner2021}, Finzi et al. \cite{finzi2020}, and Fuchs et al. \cite{fuchs2020}. Recently, Dym et al. \cite{dym2020} has proven the universality of the joined group of translations, rotations, and permutations of the SE(3)-Transformer \cite{fuchs2020}.
 
Here, we implement a custom SE(3)-Transformer and train various networks on a subset of the ANI-1x database of 5 million molecules containing four atom types (C,H,N,O). The training objective is the accurate prediction of energies and forces. The forces are predicted directly as the output of operations on the molecular graphs, unlike ANI, which predicted forces indirectly using the negative gradient of the energy predictions.

\section{Methods}

\subsection{Architecture}

We implemented a custom SE(3)-Transformer in PyTorch using the Deep Graph Library (DGL) \cite{wang2019} and made it available on Github \cite{github}. For details about the original SE(3)-Transformer implementation, refer to Fuchs et al. \cite{fuchs2020}. The differences in the this implementations include an attention block without skip connections and usage of DGL's message passing functions. As shown in Figure 1, we constructed a basic network with three SE(3)-Transformer multi-head attention layers. Each multi-head attention layer passes its input vector to each of its heads. The layer then concatenates the outputs from all the heads. The concatenated outputs are then shrunk using an equivariant linear layer and the output is returned. We trained six different SE(3)-Transformers with architectures and test performances summarized in Table 1.

The input, intermediate, and output features of the SE(3)-Transformer are concatenated vectors of so called types. Type-$\ell$ features are vectors of length $2\ell+1$ and are rotated by $(2\ell+1) \times (2\ell+1)$ Wigner-D matrices, which are representations of the group SO(3). Type-0 vectors are scalars and type-1 vectors are ordinary vectors in $\mathbb{R}^3$. The SE(3)-Transformer networks output a scalar (type-0) and a vector (type-1) for each node. The sum of all the scalars over all atoms is the predicted energy of the molecule. The vectors are the predicted force on each atom.

\begin{figure}[h!]
  \centering
  \includegraphics[scale=0.2]{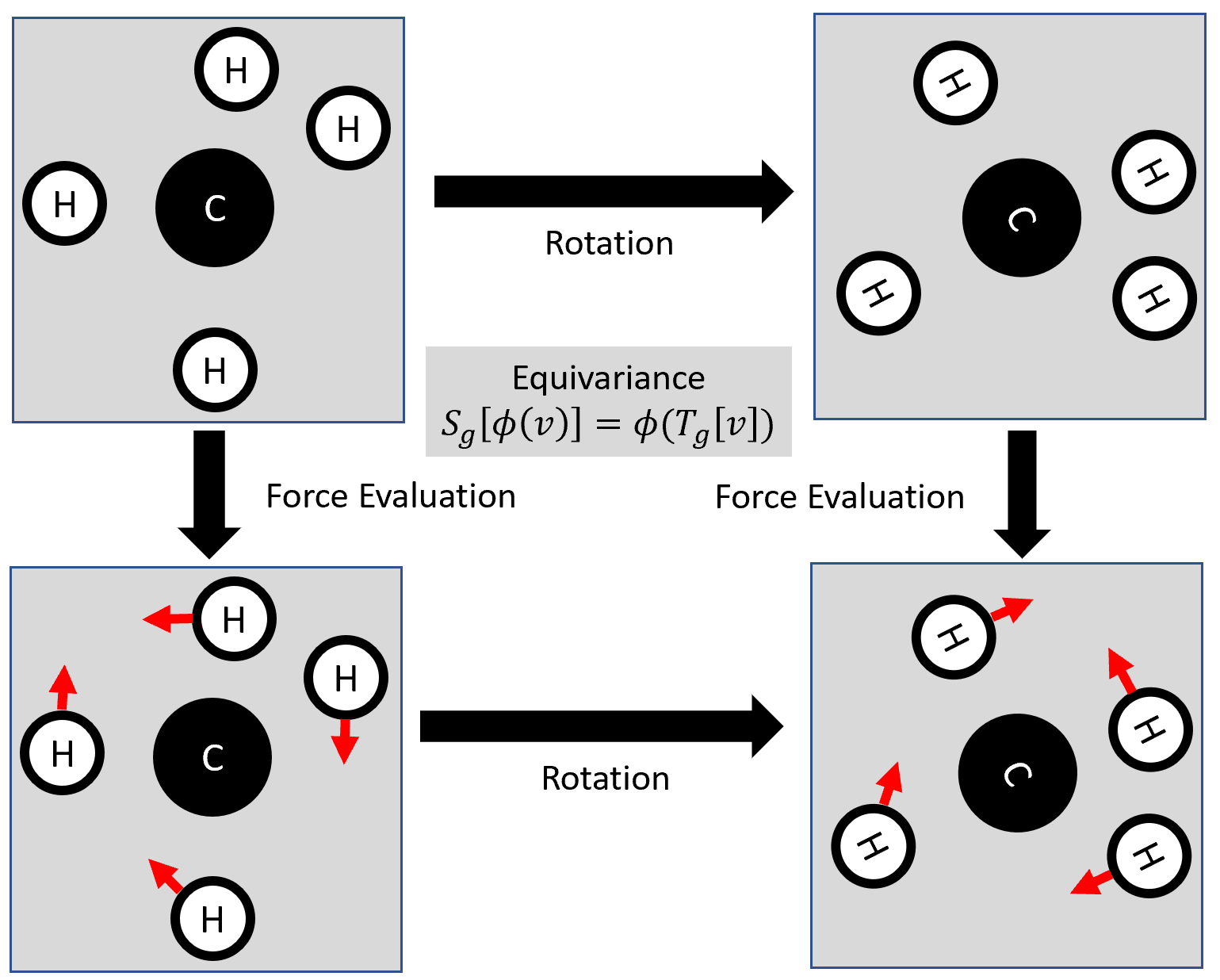}
  \hspace{0.25cm}
  \includegraphics[scale=0.17]{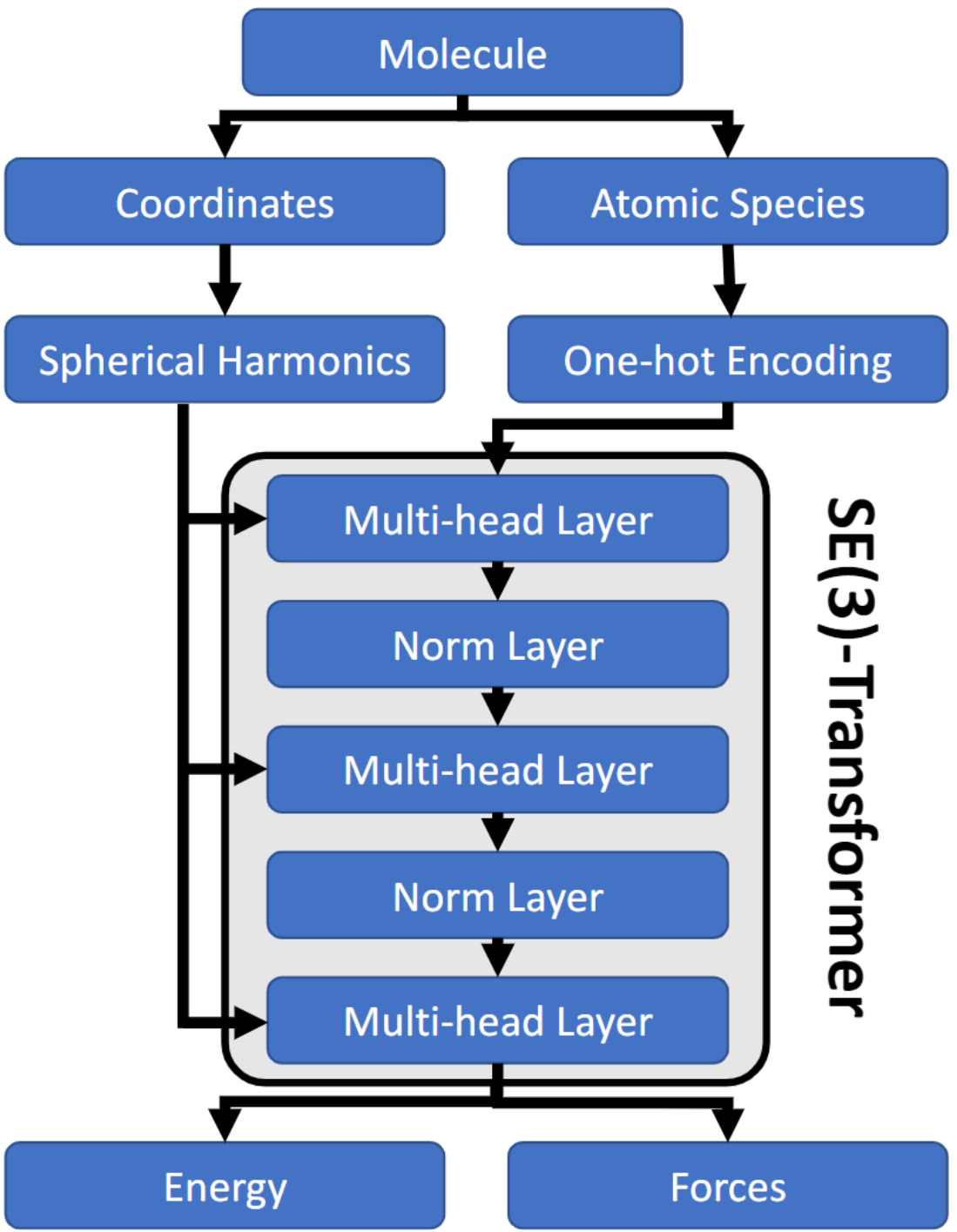}
  \caption{Left: Equivariance of rotation and force calculations visualized. Given a set of transformations $T_g:V\rightarrow V$ for elements $g\in\text{SO}(3)$, the function $\phi:V\rightarrow V$ is equivariant if it follows the relation in the figure for transformations $S_g:Y\rightarrow Y$ for all $g\in \text{SO}(3)$ and $v\in V$.
  Right: Architecture of basic SE(3)-Transformer.}
  \label{fig:my_label}
\end{figure}

\subsection{Input features}

Our networks use two categories of inputs: relative positions and atom type encodings. The relative positions are encoded as basis matrices, which are calculated using the spherical harmonics of the displacement vector between adjacent atoms and the Clebsch-Gordan coefficients, similar to Thomas et al. \cite{thomas2018} and Weiler et al. \cite{weiler2018}. The input node features for our network is a type-0, one-hot encoding of the atomic species. In other words, the input to the network is a matrix of shape $(N,4)$ where $N$ is the number of atoms in a molecule and $4$ is the number of atomic species.

\subsection{Training and Validation}

The networks were trained on the ANI-1x dataset \cite{smith2020}. This dataset consists of five million molecular conformations (containing C, H, N, and O atoms), created in Gaussian \cite{frisch2016} using the $\omega$b97x functional \cite{chai2008} and the 6-31G* Pople basis set \cite{rassolov1998}. The forces and energies were indexed using the keys ['wb97x\_dz.forces', 'wb97x\_dz.energy']. The features of the dataset are the atomic species and coordinates and the labels of the dataset are the energies and forces. The dataset was split into training, validation, and test subsets with a 90-5-5 ratio. A stratified split was used to group all conformations of a given molecule into the same subset to prevent data leakage and to test for generalizability. The losses were calculated similar to the losses of the ANI-2 \cite{devereux2020} network:

\begin{equation}
  \textbf{Loss}=\frac{1}{N}\sum_{i=1}^N\left[
  (\hat{E}_i-E_i)^2+\frac{1}{M}\sum_{j=1}^{M_i}(\hat{f}_j-f_j)^2\right]
\end{equation}

The networks were trained on 6 Quadro RTX 5000 16GB for 1 million iterations (approximately 100 epochs). The batch size per network varied to maximize GPU usage. We trained all networks with a learning rate of 1e-3 and clipped all gradients with a norm above 10.0 for 1 million iterations, which took on average about one month per network. A validation loss was calculated on the entire validation set after every 10,000 training iterations.

\subsection{Testing}

Of the 5 million conformation in ANI-1x dataset, 250 thousand were set aside as a test set. After training for one million iterations, the last ten saved snapshots (at 991,000, 992,000, …, and 1,000,000 iterations) for each architecture were selected for testing. After running inference with these models, the resulting energies and forces were averaged to yield the reported test scores.
In addition to averaging over the snapshots snapshot, we also ensembled the three best performing networks--deeper, higher, and heads--by averaging all their forces and energy predictions.

\section{Results and Discussion}

Here, we train six different architectures of SE(3)-Transformers to predict interatomic forces and energies (see Figure 2). We found that the deeper model produced the best predictions (see Table 1), with RMSE values of 2.38 kcal/mol for energy predictions and 3.32 kcal/mol/Å for force predictions. We also found that ensembling the top three architectures (deeper, higher, heads) resulted in an even better prediction with RMSE values of 2.21 kcal/mol for energy predictions and 3.18 kcal/mol/Å for force predictions. These values indicate that our models are learning and generalizing but are still larger than the approximate accuracy of DFT calculations, which are on the order of 1 kcal/mol using $\omega$b97x/6-31G* \cite{wodrich2007}.

It is not straight forward to compare this performance with ANI. When the ANI-1 network was trained on a subset of the ANI-1 dataset, it achieved an energy test RMSE of 1.45 kcal/mol on a disjoint subset of the ANI-1 dataset.\cite{smith2017-network} Unfortunately, no test RMSE for forces and energies have been reported for ANI on a ANI-1x test set. The COMP6 test set has instead been used to benchmark ANI after training on ANI-1x, with RMSEs of 3.37 kcal/mol and 5.29 kcal/mol/Å for energy and force, respectively.\cite{comp6} We did not evaluate our models on COMP6 as the molecules in the benchmark follow a distribution of graph topologies that is different from the ANI-1x dataset. Over 12\% of the molecules in the COMP6 dataset exceed the average ANI-1x molecule size by three standard deviations. Our network is optimized for small drug-like molecules and is not expected to perform well on larger molecules.

\begin{figure}[h!]
\centering
\includegraphics[scale=0.26]{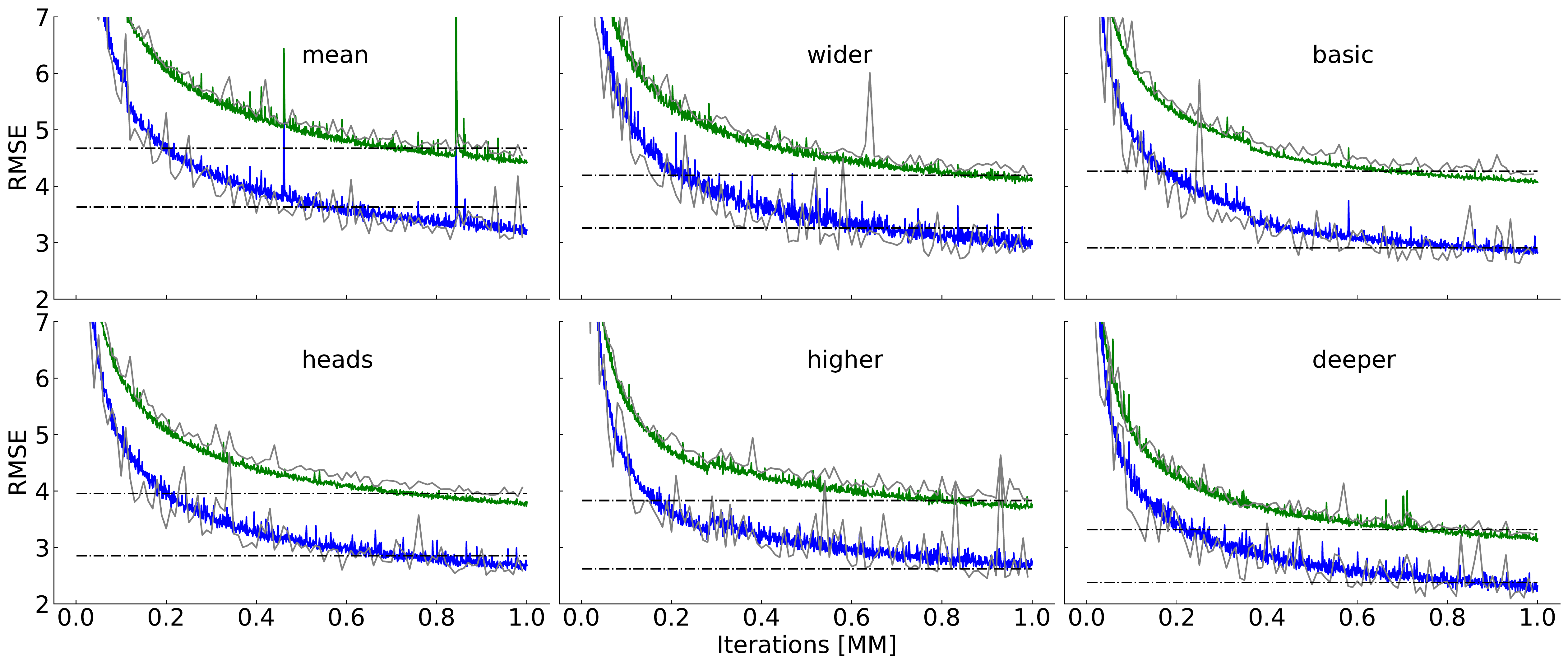}
\caption{Training, validation, and testing on six different SE(3)-Transformer architectures. The energy training curves are shown in blue and the force training curves are shown in green. Validation curves are overlaid. Test losses are shown as horizontal dotted lines.}
\end{figure}

The validation errors continued decreasing and remained close to the training errors indicating the networks never overfit to the data. Furthermore, the deeper model performed significantly better than the other models. These results suggest that an even deeper model---with additional SE(3)-Transformer layers---would likely result in further improved predictions. However, the current implementation is prohibitively slow, as it took over one month to train the deeper network. Nevertheless, a more efficient implementation, as recently proposed by Nvidia \cite{milesi2021}, would enable the necessary speed up to verify this hypothesis.

\renewcommand{\arraystretch}{1.5}
\begin{table}[h!]
  \centering
  \begin{tabular}{p{0.08\linewidth} p{0.38\linewidth} p{0.22\linewidth} p{0.22\linewidth}}
     Network & Architecture & Energy RMSE$\pm$STD [kcal/mol]& Force RMSE$\pm$STD [kcal/mol/\AA] \\
     \hline
     \hline
     \rowcolor{Gray}
     basic & 3 SE(3)-T layers, 4 heads, 4 channels, type-2 features, and sum aggregation & $2.91\pm1.12$ & $4.26\pm0.30$ \\
     mean & Mean aggregation & $3.63\pm1.40$ & $4.67\pm0.36$ \\
     \rowcolor{Gray}
     higher & Type-3 features & $2.62\pm1.38$ & $3.83\pm0.37$ \\
     wider & 8 channels & $3.26\pm1.24$ & $4.19\pm0.33$\\
     \rowcolor{Gray}
     deeper & 4 SE(3)-Transformer layers & $\textbf{2.38}\pm1.24$ & $\textbf{3.32}\pm0.33$ \\
     heads & 8 heads & $2.86\pm1.25$ & $3.96\pm0.31$\\
     \hline
     \rowcolor{Gray}
     ensemble & Averaging deeper, higher, heads & $\textbf{2.21}\pm1.47$ & $\textbf{3.18}\pm0.52$\\
  \end{tabular}
  \caption{Overview of different networks and performance on test set. In the first column are the names of the networks. The architectures of the networks are described in the second column, comparing each network to the basic network. The third and fourth columns report the energy and force performance of the networks on the test set respectively. The first item in the columns are the RMSE over the entire the dataset and the second item are the STD are the ensembling errors.}
  \label{tab:my_label}
\end{table}

The parameters of the SE(3)-Transformer networks trained here are agnostic to the atomic species provided in the inputs. Compared to the ANI algorithms, this allows rapid extension to other atomic species. Given sufficiently large datasets, such as the one used to train ANI-2, the SE(3)-Transformer could be trained to predict energies and forces for increasingly complex molecules. It is also important to point out that the networks trained here did not require handcrafted features, such as constructing atomic environment vectors. This study provides evidence that deeper SE(3)-Transformer networks can reasonably be expected to reach DFT quality energy predictions.

In the future, the effects of neighborhood definitions on the molecular graphs to achieve size independence should be explored. Deeper networks allow messages to propagate between distant neighborhoods and may result in more accurate modeling of long-range interactions in macromolecules. Once better accuracy is achieved, a trained model may be incorporated into molecular dynamics software to get DFT-accurate predictions thousands of times faster than explicit calculations. The ability to predict ab initio quality energies and forces would enable reactive simulations that overcome some of the well-known limitations of classical molecular dynamics, namely the absence of polarization effects and bond breaking/formation. 

\bibliographystyle{unsrtnat} % sort by appearance
\bibliography{endnote}

\end{document}